\documentclass{PoS}

\title{AGN variability at hard X-rays}

\ShortTitle{AGN variability at hard X-rays}

\author{\speaker{Simona Soldi}\\
        Laboratoire AIM - CNRS - CEA/DSM - Universit\'e Paris Diderot (UMR 7158), CEA Saclay, 
	DSM/IRFU/SAp, F91191 Gif-sur-Yvette, France\\
        E-mail: \email{simona.soldi@cea.fr}
	}

\author{Gabriele Ponti\\
	School of Physics and Astronomy, University of Southampton, Highfield, 
	Southampton S017 1BJ, UK\\
        E-mail: \email{ponti@apc.univ-paris7.fr}
	}
\author{Volker Beckmann\\
	APC, Centre Fran\c{c}ois Arago, IN2P3/CNRS - Universit\'e Paris Diderot, 
	10 rue Alice Domon et L\'eonie Duquet, 75205 Paris Cedex 13, France\\
        E-mail: \email{beckmann@apc.univ-paris7.fr}
	}
\author{Piotr Lubi\'nski\\	        
        Nicolaus Copernicus Astronomical Center, Polish Academy of Sciences
	ul. Rabianska 8, 87-100 Torun, Poland\\
        E-mail: \email{lubinski@ncac.torun.pl}
	}

\abstract{We present preliminary results on the variability properties of AGN above 20~keV in order to show the
	  potential of the \textit{INTEGRAL} IBIS/ISGRI and \textit{Swift}/BAT instruments for hard X-ray timing analysis of AGN.
	  The 15--50~keV light curves of 36 AGN observed by BAT during 5 years show significantly larger variations 
	  when the blazar population is considered (average normalized excess variance $\langle \sigma^2_{\rm NXS} \rangle = 0.25$)
	  with respect to the Seyfert one ($\langle \sigma^2_{\rm NXS} \rangle = 0.09$).
	  The hard X-ray luminosity is found to be anti-correlated to the variability amplitude in Seyfert galaxies and  
	  correlated to the black hole mass, confirming previous findings obtained with different AGN hard X-ray samples.
	  We also present results on the Seyfert 1 galaxy IC~4329A, as an example of spectral variability study with 
	  \textit{INTEGRAL}/ISGRI data. The position of the high-energy cut-off of this source is found to have
	  varied during the \textit{INTEGRAL} observations, pointing to a change of temperature of the Comptonising medium.
	  For several bright Seyfert galaxies, a considerable amount of \textit{INTEGRAL} data have already been
	  accumulated and are publicly available, allowing detailed spectral variability studies at hard X-rays.
	  
         }

\FullConference{The Extreme sky: Sampling the Universe above 10 keV\\
                 October 13-17 2009\\
                 Otranto (Lecce) Italy}

\begin{document}

\section{Introduction}

Variability at all observed wavelengths is a distinctive characteristic of objects hosting an accreting black hole,
in particular AGN.
The study of the timing properties of AGN can provide important information about the structure, 
the physics and the dynamics of the radiating source.
In the last years much progress has been made to characterize the AGN variability
in the 2--20 keV X-ray range, in particular since the launch of \textit{RXTE} in 1995.
On the other hand, the global variability properties of AGN at hardest X-rays above 20 keV have 
been poorly studied in the past. 
Some studies of the spectral variability of AGN during different flux states were carried out
with \textit{CGRO}/OSSE and \textit{BeppoSAX}/PDS (e.g. \cite{petrucci00}), while a long term monitoring at hard X-rays 
was not possible due to the relatively small field of view of these instruments and the observation strategy of these satellites.
The only exception was the BATSE instrument on board of \textit{CGRO}, that detected a handful of AGN
using the Earth occultation technique \cite{harmon04,soldi08}.

\textit{INTEGRAL} IBIS/ISGRI \cite{winkler03,lebrun03} and \textit{Swift}/BAT \cite{gehrels04} offer now a unique opportunity to observe a large number of 
AGN on different time scales in the hard X-ray band above 20 keV. 
The first instrument is more suited for investigating the spectral variations in bright and well monitored AGN, 
as already shown, for example, for NGC~4151 \cite{lubinski10} and MCG$-$05$-$23$-$016 \cite{beckmann08}.
On the other hand, the BAT provides a database to study the timing properties 
of a large AGN sample. 

The results obtained from the first 9 months of \textit{Swift}/BAT observations have been reported for 44 AGN detected
at more than $10\sigma$ over this time period \cite{beckmann07}. 
The blazars had been found to show the highest variability, and 30\% of the Seyfert galaxies in the sample showed variability 
of $> 10\%$ with respect to their average flux. 
A general trend of increasing variability with absorption had been detected that could be ascribed to other two relations, i.e.
the anti-correlation between absorption and luminosity (observed by different X-ray surveys \cite{ueda03,beckmann09}) 
and between variability and luminosity, observed also in the BAT sample.

\begin{figure}
\includegraphics[width=.45\textwidth,angle=90]{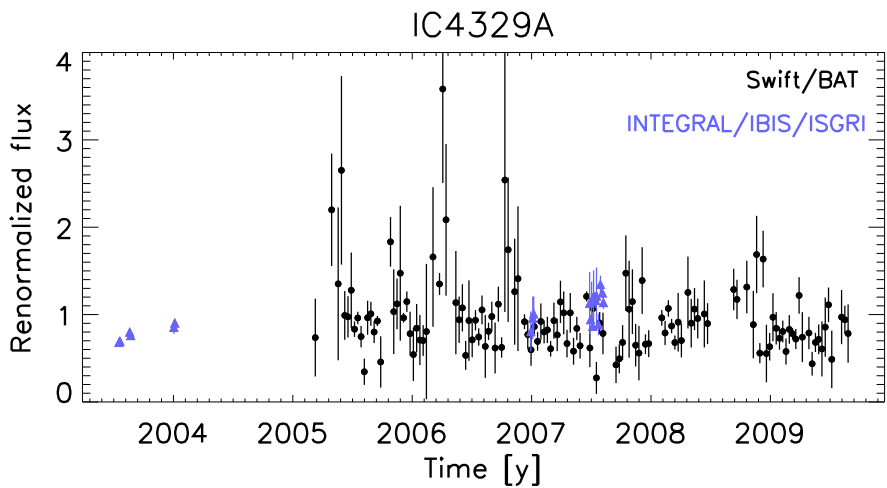} 
\caption{15--50 keV \textit{Swift}/BAT (black circles) and 20--60 keV \textit{INTEGRAL}/ISGRI (blue triangles) light curves of the Seyfert 1 galaxy 
IC~4329A. For better clarity, the light curves are rebinned to 10 and 3 days, respectively. } 
\label{lc_IC4329A} 
\end{figure}

About 6 years of public \textit{INTEGRAL} data and 5 years of \textit{Swift} data are currently available, providing the possibility
to extend this kind of studies to a much larger AGN sample and with better statistic.
In Fig.~\ref{lc_IC4329A} an example of BAT and ISGRI hard X-ray light curves for the bright Seyfert 1 galaxy IC~4329A is shown.

\section{\textit{Swift}/BAT long-term monitoring}
Since November 2004, the instrument BAT \cite{barthelmy05} on board the \textit{Swift} satellite has been observing the sky
in the 15--195 keV energy range and, thanks to its large field of view of $\sim 1.4 \rm \, sr$ and to \textit{Swift}'s observing strategy, 
it has been monitoring a large number of hard X-ray sources \cite{tueller09}.
Daily light curves in the 15--50 keV band are provided on the \textit{Swift}/BAT hard X-ray transient monitoring 
pages\footnote{\emph{http://swift.gsfc.nasa.gov/docs/swift/results/transients}} for 281 AGN. 
We have downloaded the BAT light curves for these AGN, covering the time from the beginning of the mission up to August 23, 2009.
The light curves have been rebinned to 3-day bins and we selected the 36 AGN for which the average of the detection significance
of each point of the light curve is larger than 3$\sigma$. This limit has been chosen after checking that
the light curves of random positions in the sky (i.e. background) provided by the BAT team have an average
significance of $\leq 2.4 \sigma$ when rebinned to 3-days.
Among these 36 AGN, there are 18 blazars, 3 Seyfert 1 galaxies, 3 intermediate type Seyfert, 8 Seyfert 2 and 4 radio galaxies.

\begin{figure}
\includegraphics[width=.35\textwidth,angle=90]{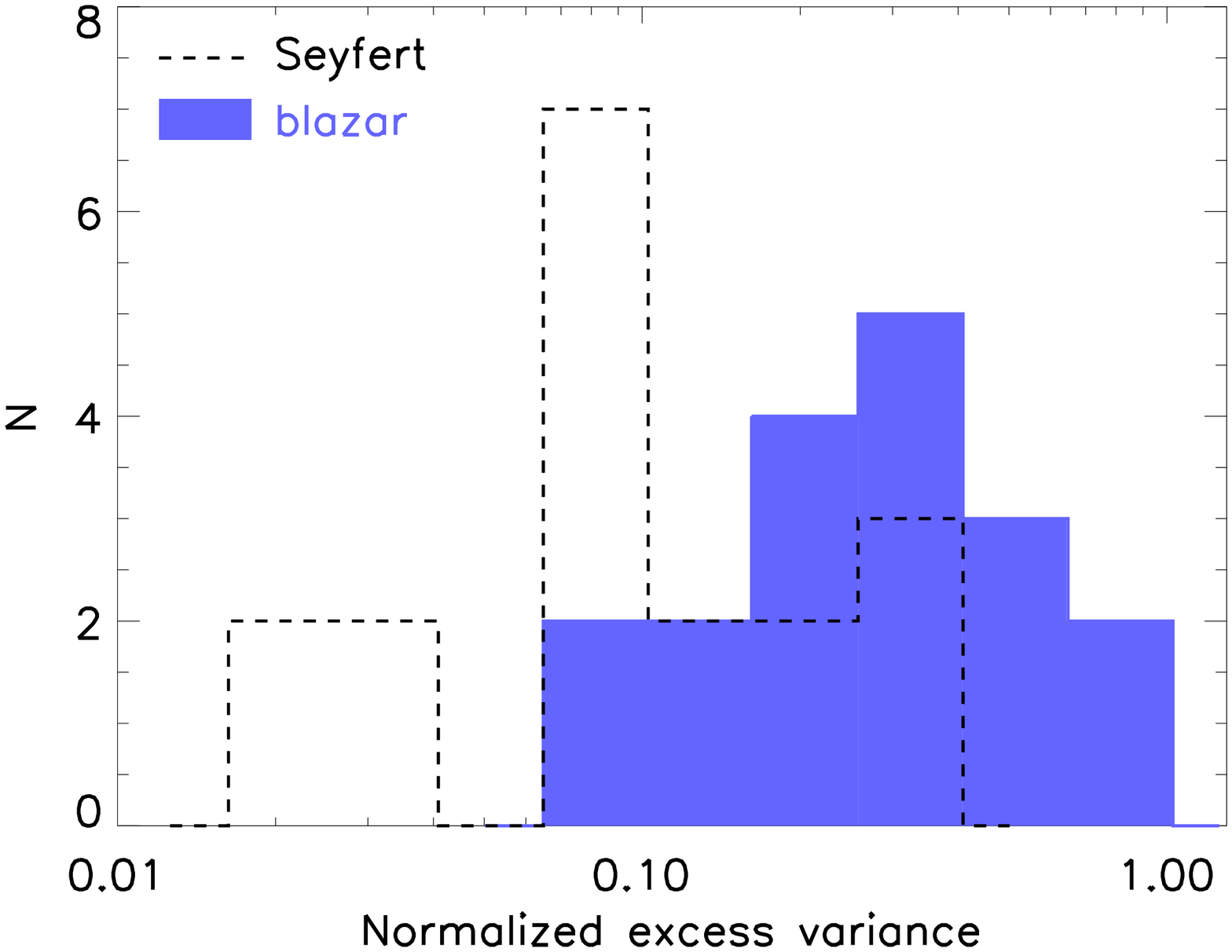} 
\includegraphics[width=.35\textwidth,angle=90]{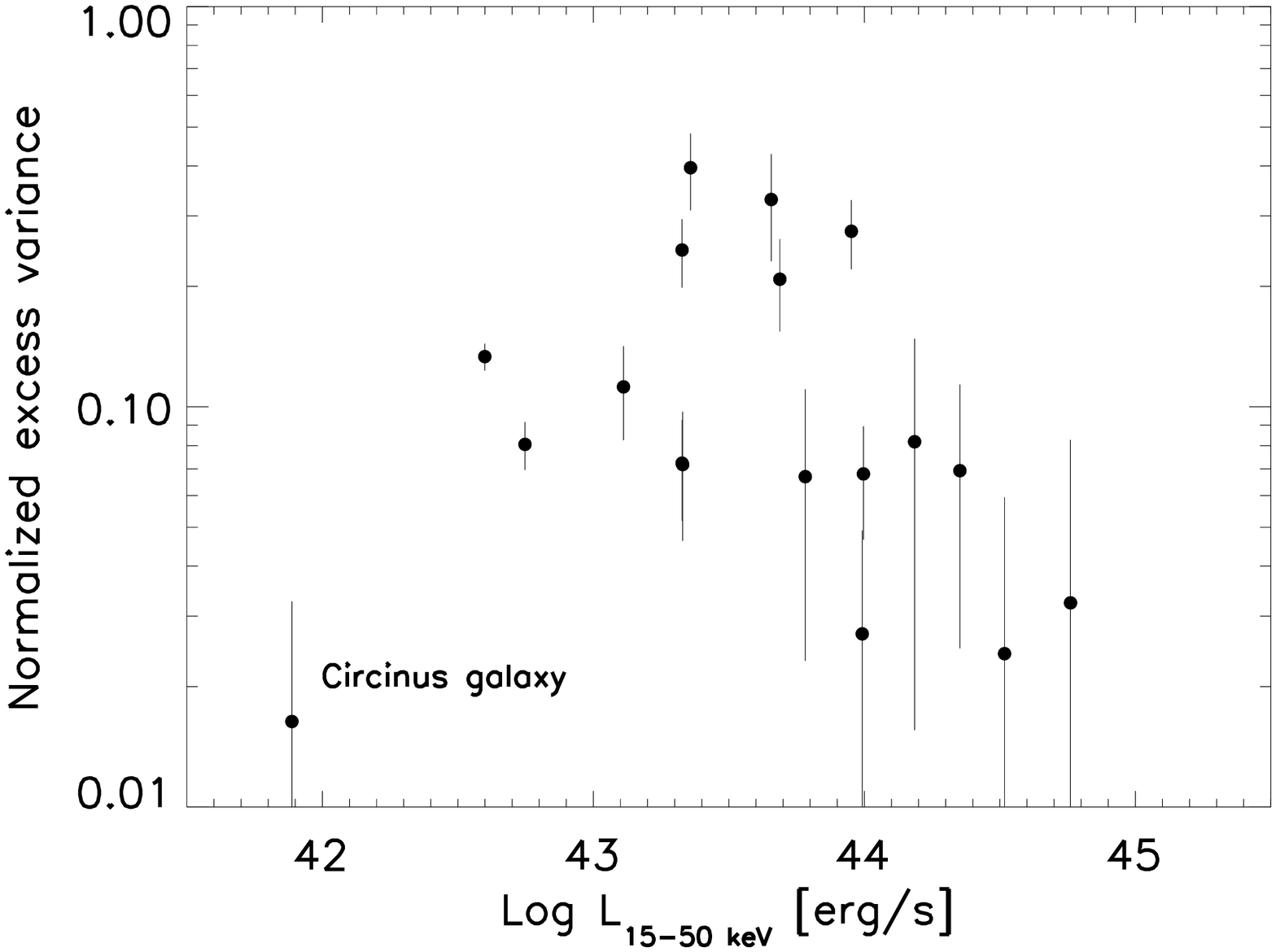} 
\caption{\textit{Left:} Histograms of the 15--50 keV normalized excess variance computed for 36 AGN observed by 
	\textit{Swift}/BAT (Seyfert and radio galaxies in black, blazars in blue).
	\textit{Right:} Normalized excess variance versus 15--50 keV luminosity for the Seyfert and radio galaxies 
	in our BAT AGN sample.
	}
\label{histo} 
\end{figure}

We use this sample to study the hard X-ray variability of AGN, applying some basic variability estimators, as the normalized excess variance $\sigma^2_{\rm NXS}$
\cite{vaughan03,ponti04}. 
The 18 Seyfert and radio galaxies show an average (in logarithmic space) amplitude of variations $\langle \sigma^2_{\rm NXS} \rangle = 0.09$, whereas the 18 blazars 
show on average larger variations, $\langle \sigma^2_{\rm NXS} \rangle = 0.25$ (left panel in Fig.~\ref{histo}). 
We find a 1\% KS-test probability that the two samples are drawn from the same distribution. 
The larger variability of blazars at hard X-rays is expected due the nature
of their emission in this energy range, usually believed to be produced in relativistic jets (e.g. \cite{pian08}).
According to our preliminary results, there does not seem to be any correlation between the normalized excess variance and the black 
hole mass nor the Eddington ratio for the AGN in this sample, whereas we confirm the trend observed 
in Seyfert and radio galaxies of less luminous objects being more variable (right panel in Fig.~\ref{histo}) already reported by Beckmann et al. (2007).
For this relation, we find a probability of chance correlation of 2.5\%, when the Circinus galaxy is excluded. 
This peculiar object is a Compton thick Seyfert 2 galaxy with high reflection through a thick torus \cite{yang09}, which could explain the low variability observed.

\begin{figure}[!b] 
\begin{center}
\includegraphics[width=.4\textwidth,angle=90]{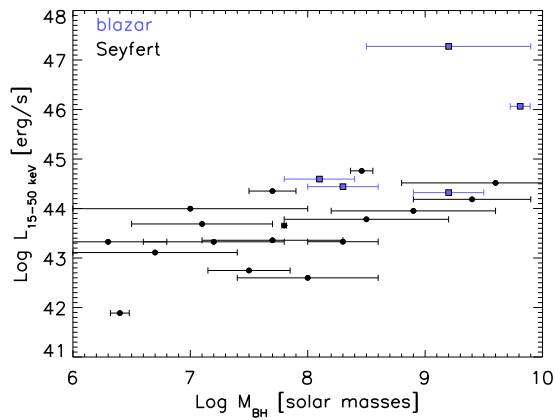} 
\caption{Correlation between the 15--50 keV luminosity and the central black hole mass for our BAT AGN sample. 
	Seyfert and radio galaxies are indicated with black circles, blazars with blue squares.} 
\label{MBH_Lx} 
\end{center}
\end{figure}

We also confirm the correlation already found for other X-ray selected AGN samples 
\cite{beckmann09,bianchi09} between the luminosity (averaged over the 5 years of Swift observations) 
and the black hole mass for the 22 AGN with measured masses (Fig.~\ref{MBH_Lx}).
We obtain a probability of chance correlation of 0.03\% (1\% when only the 17 Seyfert galaxies are considered) and a relation of the form
$L_X \propto M_{\rm BH}^{(0.75 \pm 0.18)}$ ($L_X \propto M_{\rm BH}^{(0.5 \pm 0.2)}$ for Seyfert galaxies only). 
A proportionality with index lower than 1, as found here, indicates that the more massive objects have either a lower X-ray efficiency or a 
lower accretion rate than less massive objects. The latter (i.e. sub-Eddington luminosities for the most luminous, high-mass quasars) 
has been recently observed for a large sample of quasars up to redshift $z = 2$ from the Sloan Digital Sky Survey \cite{steinhardt09}.

\section{Spectral variability: the example of IC~4329A}
\begin{figure}
\hspace{-0.8cm}
\includegraphics[width=.43\textwidth,angle=90]{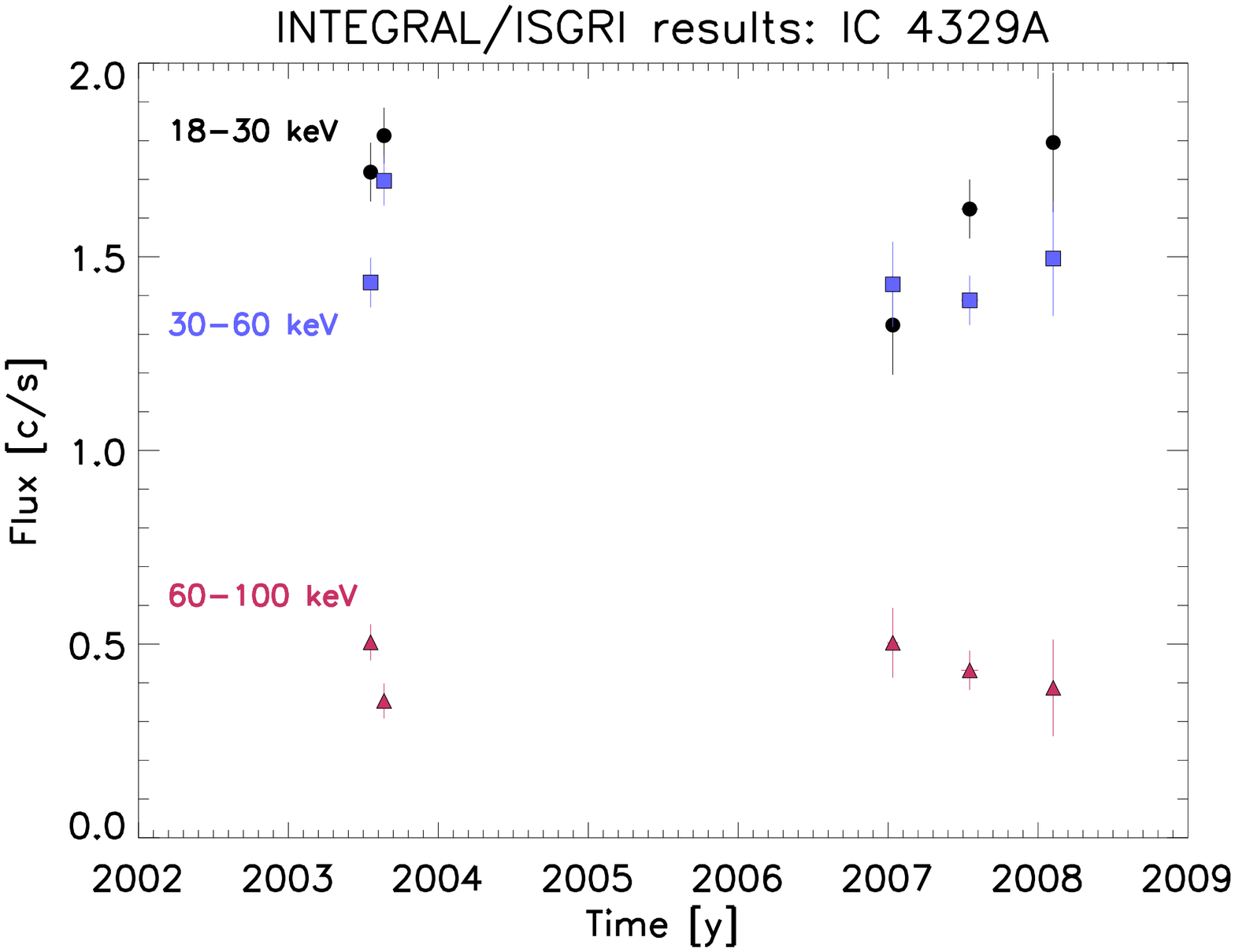} 
\includegraphics[width=.53\textwidth,angle=90]{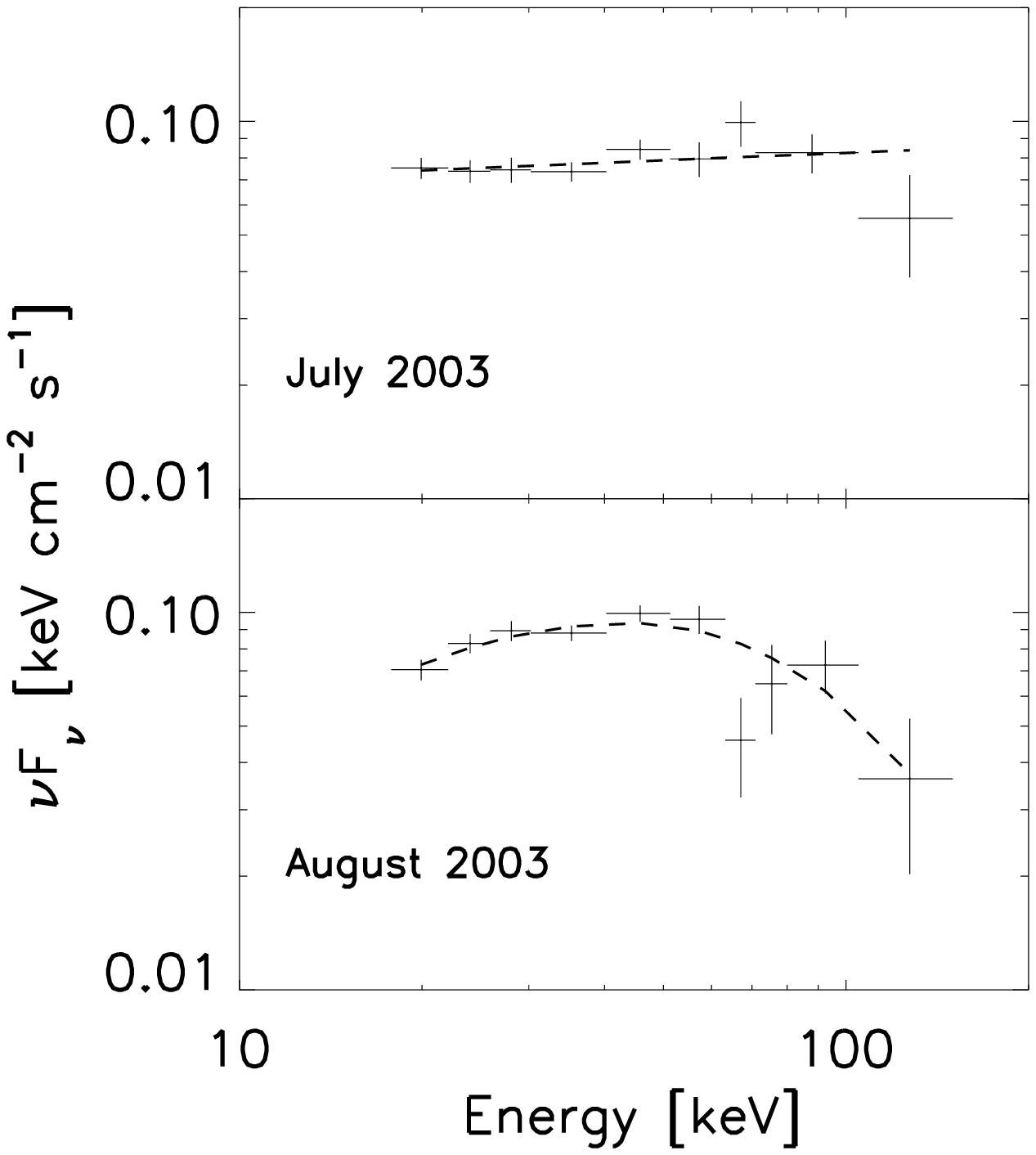} 
\caption{\textit{Left:} \textit{INTEGRAL}/ISGRI light curve of IC~4329A in three energy bands. Each data point
	corresponds to an exposure time from 50 to 200 ks accumulated within 2 to 40 days.
	\textit{Right:} ISGRI spectra of IC~4329A extracted from the \textit{INTEGRAL} observations of July 2003 (top)
	and August 2003 (bottom). The spectra are fitted with a simple power law (top) and with an exponentially
	cut-off power law, with cut-off at $39 \pm 17 \rm \, keV$ (bottom).
	}
\label{spe} 
\end{figure}

For the brightest AGN, accurate spectral variability studies in the hard X-ray band can be carried out with \textit{INTEGRAL} data 
(e.g. \cite{lubinski10,beckmann08}). As an example, the spectral variability of the bright Seyfert 1 galaxy IC~4329A has been investigated
in the 20--100 keV band.
We have analysed all the \textit{INTEGRAL} IBIS/ISGRI data available at the time of our study, which includes data for an effective exposure time of about 
620~ks collected from the beginning of the mission up to February 12, 2008. The data have been analysed using the version 8 of the Offline
Scientific Analysis Software (OSA).
The data can be separated into 5 different periods, spanning from 50 to 200 ks of effective exposure time each, accumulated within
2 to 40 days. 
The spectra for each period have been extracted using the standard OSA spectral extraction and 3\% systematics have been added.
In the left panel of Figure~\ref{spe}, the uncorrelated variations of the 18--60 and 60--100 keV fluxes suggest a spectral 
change of the source during the 6 years of \textit{INTEGRAL} observations. Indeed, a high-energy cut-off at $E_{\rm c} = 39 \pm 17 \rm \, keV$, 
signature of thermal Comptonisation processes, can be found in the August 2003 ISGRI spectrum, whereas no sign of curvature was 
detectable in July 2003 (right panel of Fig.~\ref{spe}).
During the same period (August 2003) the broad component of the 6.4~keV Fe K line has been found to show a hint of variability
on hour time scale \cite{demarco09}. 
This hard X-ray spectral change could indicate an increase of the electron plasma temperature, moving the cut-off towards higher energies, where it was not detectable
by ISGRI in the July 2003 observation. 
Adding data below 20~keV will allow us to fit the combined spectrum with a more physical Comptonisation model, and therefore, to more precisely determine
the temperature of the Comptonising plasma and constrain the presence and strength of the Compton reflection hump,
already observed for this source in the past by \textit{BeppoSAX} \cite{dadina07}.

\vspace{-0.3cm}
\section{Conclusions}
We have presented preliminary results of a study investigating the variability properties of AGN at hard X-rays.
The 15--50 keV light curves of a sample of 36 AGN out of the 281 monitored by \textit{Swift}/BAT have been analysed.
We found a significantly larger variability of the blazar population compared to the Seyfert one, as expected
considering the beamed nature of the high-energy emission of blazars.
We confirm the trend of increasing hard X-ray variability with decreasing luminosity for the Seyfert sample,
and the correlation between the hard X-ray luminosity and the black hole mass for the 22 AGN with known masses \cite{beckmann07,beckmann09}.
The latter relation is consistent with more massive black holes accreting at lower Eddington rates than the less massive ones.

As an example of spectral variability studies with \textit{INTEGRAL}, we have analysed IBIS/ISGRI
data spanning more than 5 years of the bright Seyfert 1 galaxy IC~4329A. Spectral variations have been observed, with a high-energy cut-off being 
detected in August 2003 whereas no curvature was visible in the July 2003 spectrum.
Further analysis of the \textit{INTEGRAL} spectra, including also the JEM-X data below 20~keV will allow
to further investigate this possible change in the plasma electron temperature and the presence of a reflection
component.

Even though only preliminary, the results presented here show the potential of \textit{INTEGRAL}/ISGRI and 
\textit{Swift}/BAT data to study AGN variability above 20~keV.
The large amount of \textit{INTEGRAL} data currently available will enable us to extend this kind of studies 
to about 20 Seyfert galaxies, among which MCG~$+$08$-$11$-$11, NGC~4593, NGC~4388, Mrk~509, NGC~2110, as already
done for other AGN \cite{beckmann08,lubinski10}. In the meantime, the BAT monitoring is continuing,
providing us with the best sampled hard X-ray light curves for an increasing sample of AGN.

\vspace{-0.3cm}
\section*{Acknowledgments}
\vspace{-0.3cm}
\small{ 
S.S. acknowledges the Centre National d'Etudes Spatiales (CNES) for financial support.
Part of the present work is based on observations with \textit{INTEGRAL}, an ESA project with
instruments and science data centre funded by ESA member states with the participation of Russia and the USA.
}

\end{document}